# Poisson's ratio effects on the mechanics of auxetic nanobeams

## S. Faroughi[a] and M. Shaat[b,c,*]

[a] *Faculty of Mechanical Engineering, Urmia University of Technology, Urmia, Iran*

[b] *Engineering and Manufacturing Technologies Department, DACC, New Mexico State University, Las Cruces, NM 88003, USA*

[c] *Mechanical Engineering Department, Zagazig University, Zagazig 44511, Egypt*

## Abstract

Poisson's ratio is an important mechanical property that reveals the deformation patterns of materials. A positive Poisson's ratio is a feature of the majority of materials. Some materials, however, display "auxetic" behaviors (i.e. possess negative Poisson's ratios). Indeed, auxetic and non-auxetic materials display different deformation mechanisms. Revealing these differences and their effects on the mechanics of these materials is of a significant importance.

In this study, effects of Poisson's ratio on the mechanics of auxetic and non-auxetic nanobeams are revealed. A parametric study is provided on effects of Poisson's ratio on the static bending and free vibration behaviors of auxetic nanobeams. The general nonlocal theory is employed to model the nonlocal effects. Unlike Eringen's nonlocal theory, the general nonlocal theory uses different attenuation functions for the longitudinal and lateral strains. This theory can reveal the Poisson's ratio-nonlocal coupling effects on the mechanics of nanomaterials. The obtained results showed that Poisson's ratio is an essential parameter for determining mechanical behaviors of nanobeams. It is revealed that auxetic and non-auxetic nanobeams may reflect softening or hardening behaviors depending on the ratio of the nonlocal fields of the beam's longitudinal and lateral strains.

## 1. Introduction

Poisson's ratio is a vital measure for elastic-deformation of materials. Poisson's ratio, $\nu$, is explained as the ratio of the lateral contraction in a solid material to its longitudinal extension due to an axial tension (Lakes, 1993). The significant importance of Poisson's ratio is that it provides a vision on the structural behavior of materials. For instance, Poisson's ratio can be considered as a measure of the compressibility of the material (Greaves et al., 2011). Liquids and rubbers possess $\nu \to 0.5$ while glasses and minerals

-------------------------------------------------

*Corresponding author: Tel.: +15756215929 (M. Shaat)

*E-mail address:* sh.farughi@uut.ac.ir (S. Faroughi) & shaat@nmsu.edu; shaatscience@yahoo.com (M. Shaat).

possess $\nu \to 0$. Network structures (*e.g.* metamaterials) exhibit negative Poisson's ratio, $\nu < 0$. Generally, the Poisson's ratio of a material ranges from -1 to 0.5.

Materials with negative Poisson's ratio known as '*auxetic materials*'. Negative Poisson's ratios were revealed in various materials under special conditions. For example, Baughman et al. (1998) demonstrated that 69% of cubic metallic crystals possess negative Poisson's ratios. Milstein and Huang (1979) revealed that FCC crystals exhibit negative Poisson's ratios along the transverse direction [1$\bar{1}$0] to a [110] uniaxial load. Using Molecular Dynamics (MD) simulations, HO et al. (2014) revealed the negative Poisson's ratios of metallic nanoplates under finite strains. Moreover, MD revealed the dependency of the Poisson's ratio of single crystalline nanobeams on the size and crystallographic orientation (Ahadi and Melin, 2016); thus, a nanobeams loaded in the [110]-direction may exhibit negative Poisson's ratios. Furthermore, the negative Poisson's ratio (*or, in other words, the auxetic behavior of materials*) was revealed in cellular solids, crystalline and amorphous materials, composites, and metamaterials (Lakes, 1993; Greaves et al., 2011).

Some studies were conducted to reveal Poisson's ratio effects on the mechanics of structural materials. For example, it was revealed that bars or plates deform into saddle shapes due to a positive Poisson's ratio. Negative Poisson's ratios, on the other hand, cause a convex shape in bended plates (Lakes, 2017). Moreover, it was demonstrated that Poisson's ratio affects the wave speed, strain and toughness of materials. Thus, longitudinal waves were obtained propagating faster than shear waves if Poisson's ratio is large and vice versa if Poisson's ratio is small (Lakes, 2017). By mixing single-walled carbon nanotubes with multi-walled carbon nanotubes, Hall et al. (2008) demonstrated that the in-plane Poisson's ratio of the nanotube can be converted from positive to negative which increases the tube's toughness, strength, and modulus. In some studies, effects of the Poisson's ratio on elastic fields are discussed.

In a few studies, the mechanics of auxetic beams were investigated. For instance, the buckling behavior of anisotropic-auxetic beams were investigated in (Heyliger, 2015). Hadjigeorgiou and Stavroulakis (2004) demonstrated the applicability of auxetic cantilever beams for smart structures. Moreover, the shear deformation behavior of auxetic beams was discussed by Lim (2015). Ruzzene and Scarpa (2003) studied the wave propagation in composite beams with auxetic cores. Kumar and Chinthapenta (2014) studied the blast resistance of sandwich beams with an auxetic core. Schenk et al. (2014) studied the dynamic behavior of sandwich beams with stacked-folded auxetic cores subject to static compression and impulsive loadings. These studies revealed nontraditional mechanics for auxetic beams. Indeed, more studies should be conducted to reveal these nontraditional behaviors of auxetic beams.

In this study, the effects of the Poisson's ratio on the mechanics of auxetic nanobeams are revealed. Here, in order to account for the effects of the nonlocal-long range interatomic interactions, the general nonlocal theory developed by Shaat (Shaat and Abdelkefi, (2017); Shaat (2017)) is employed. Unlike Eringen's nonlocal theory (Eringen, 1983), the general nonlocal theory considers two distinct nonlocal fields for the longitudinal strain and the transverse strain. Eringen developed his differential nonlocal theory in 1983 assuming the same nonlocal fields for both the longitudinal and transverse strains. Thus, Eringen's nonlocal theory cannot reveal the Poisson's ratio-nonlocal field coupling effects on the material's mechanics. The general nonlocal theory, on the other hand, can reveal the Poisson's ratio effects on the nonlocal residuals in materials. To demonstrate the effects of the Poisson's ratio on the mechanics of auxetic nanobeams, the static bending and frequencies of simple supported beams are determined. The iterative nonlocal residual approach (Shaat, 2015) is employed to obtain the nonlocal fields of the simple supported auxetic nanobeams. Thus, the variations of the deflection and frequencies of simple supported auxetic beams as functions of the Poisson's ratio are depicted.

## 2. Auxetic nonlocal beams: Simple supported beams

Eringen (1972-a; 1972-b; 2002) proposed the nonlocal theory of linear elasticity in which the constitutive equations depend on only one attenuation function (or differential operator). Based on Eringen's nonlocal theory, research studies have been performed on the mechanics of nanobeams (Peddieson et al., 2003; Lu et al., 2006; Thai, 2012; Schenk et al., 2014; Reddy et al., 2014; Eltaher et al., 2016; 2013; Tuna and Kirca, 2016; Fernández-Sáez et al., 2016), nanoplates (Wang et al., 2010; Narendar and Gopalakrishnan, 2012; Zhang et al., 2014; Faroughi et al., 2017), and nanoshells (Wang and Varadan, 2007; Hu, et al. 2008, Ghavanloo and Fazelzadeh, 2013; Ansari et al., 2017).

Recently, Shaat (Shaat and Abdelkefi (2017); Shaat (2017)) demonstrated that Eringen's nonlocal model shows some limitations when applied for materials exhibiting different nonlocal behaviors for their longitudinal and transverse strains. For such materials, it was revealed that Eringen's nonlocal theory cannot simultaneously fit their experimental longitudinal and transverse acoustic dispersions. To enhance Eringen's nonlocal theory, Shaat (2017) proposed the general nonlocal theory which introduces different attenuation functions for longitudinal and transverse strains of the material. The general nonlocal theory allows for the simultaneous fitting of both longitudinal and transverse acoustic dispersions of various materials (Shaat & Abdelkefi, 2017; Shaat, 2017).

In this study, the general nonlocal theory is employed to reveal Poisson's ratio effects on the mechanics of auxetic beams. Unlike the conventional Eringen's nonlocal model, it is revealed that the general nonlocal theory models Poisson's ratio effects on nanobeams.

## 2.1. Equation of motion

Consider an isotropic beam with $x-$axis is its longitudinal axis and $y$ and $z-$axes are its transverse directions. By modeling the beam as an Euler-Bernoulli beam, its longitudinal strain is written as follows:

$$\varepsilon_{xx} = -z\nabla^2 w(x) \tag{1}$$

where $w(x)$ is the beam's deflection. $\nabla^2 = \partial^2/\partial x^2$ is a gradient operator.

For the considered isotropic beam, the transverse strains can be determined as follows:

$$\varepsilon_{yy} = \varepsilon_{zz} = -\nu\varepsilon_{xx} \tag{2}$$

where $\nu$ is the Poisson's ratio.

According to the general nonlocal theory (Shaat and Abdelkefi, 2017; Shaat, 2017), the non-zero constitutive equations of the beam can be written as follows:

$$(1 - \epsilon_1\nabla^2)(1 - \epsilon_2\nabla^2)t_{xx} = \frac{E}{1+\nu}\left[\frac{\nu}{(1-2\nu)}(1 - \epsilon_2\nabla^2) + (1 - \epsilon_1\nabla^2)\right]\varepsilon_{xx}$$

$$t_{yy} = t_{zz} = -\nu t_{xx} \tag{3}$$

where $t_{ij}$ are nonlocal stresses. $\epsilon_1$ and $\epsilon_2$ are two distinct nonlocal parameters. $E$ denotes the beam's Young's modulus.

Shaat (2015) developed the iterative nonlocal residual approach which allows for solving nonlocal field problems by employing local boundary conditions. In the context of this approach, the field equation is solved in a local-like field with a nonlocal-like residual field. This nonlocal residual field is iteratively formed and utilized to correct the field equation for the nonlocal residuals. Thus, with the iterations, the derived local-like fields converge to the exact nonlocal field of the original nonlocal field problem.

In the context of Shaat's iterative-nonlocal residual approach, a nonlocal stress residual $\tau_{xx}$ is formed as the difference between the local stress ($\sigma_{xx}$) and the nonlocal stress ($t_{xx}$) of a previous iteration, $k-1$, (Shaat, 2015):

$$\tau_{xx}^{(k)} = \sigma_{xx}^{(k-1)} - t_{xx}^{(k-1)} \tag{4}$$

Accordingly, the equation of motion of Euler-Bernoulli nonlocal beams can be written in the form:

$$\left(E^*\frac{bh^3}{12}\right)\nabla^4 w(x,t)^{(k)} + (\rho bh)\ddot{w}(x,t)^{(k)} = P(x,t) + F(x,t)^{(k)}$$

with

$$E^* = \frac{E}{1+\nu}\left(1 + \frac{\nu}{1-2\nu}\right) \tag{5}$$

where $b$ and $h$ are, respectively, the beam's breadth and thickness. $\rho$ is the beam's mass density. $P(x,t)$ is the applied transverse load.

It is should be noted that the equation of motion (5) is written for an iteration, $k$. The last term in the right-hand-side of equation (5) is a fictitious force which is formed depending on the nonlocal residual field, $F(x,t)^{(k)}$. In the first iteration, this fictitious force is zero. Then, with the iterations, this force grows up accumulating the nonlocal residual field in the beam. This fictitious force can be written in the form (Shaat, 2015):

$$F(x,t)^{(k)} = E^*\nabla^4 w_c(x,t)^{(k)} \tag{6}$$

where $w_c(x,t)$ is a deflection-correction field for the nonlocal residual.

By multiplying equation (4) by $(1-\epsilon_1\nabla^2)(1-\epsilon_2\nabla^2)$, the following relation between the deflection-correction field ($w_c$) and the deflection ($w$) can be derived:

$$\begin{aligned}&\left(1-\epsilon_1\nabla^2\right)\left(1-\epsilon_2\nabla^2\right)\nabla^2 w_c(x,t)^{(k)}\\&=\left\{\left(1-\epsilon_1\nabla^2\right)\left(1-\epsilon_2\nabla^2\right)-\left(\frac{\nu}{1-\nu}\right)\left(1-\epsilon_2\nabla^2\right)-\left(\frac{1-2\nu}{1-\nu}\right)\left(1-\epsilon_1\nabla^2\right)\right\}\nabla^2 w(x,t)^{(k-1)}\end{aligned} \tag{7}$$

Equation (7) can be simplified according to Eringen's nonlocal theory by setting $\epsilon_1 = \epsilon_2 = \xi^2$, as follows:

$$(1-\xi^2\nabla^2)\nabla^2 w_c(x,t)^{(k)} = -\xi^2\nabla^4 w(x,t)^{(k-1)} \tag{8}$$

where $\xi^2$ is the single-nonlocal parameter that is used in Eringen's nonlocal theory.

By comparing equations (7) and (8), it is clear that the general nonlocal theory outperforms Eringen's nonlocal theory in revealing Poisson's ratio effects.

## 2.2. Static bending of simple supported beams

In the context of the iterative nonlocal residual approach, the nonlocal field equation is converted into an equivalent local-like field equation. This permits implementing the local boundary conditions. Consequently, for a simple supported Euler-Bernoulli beam, the deflection, $w(x)$, can be written as follows:

$$w(x,t) = W\sin\left(\frac{\pi}{L}x\right) \tag{9}$$

where $L$ is the beam length. $W$ denotes the amplitude of the beam deflection. According to Shaat's iterative nonlocal residual approach, the deflection-correction field, $w_c(x)$, is considered having the same form as the beam deflection, as follows:

$$w_c(x,t) = W_c\sin\left(\frac{\pi}{L}x\right) \tag{10}$$

where $W_c$ is the amplitude of the beam deflection-correction field.

The equation governing the static bending of Euler-Bernoulli nonlocal beams according to the iterative nonlocal residual approach can be obtained by substituting equations (9) and (10) into the foregoing equations, eliminating time dependent terms from equation (5), and integrating the result over the beam length, as follows:

$$W^{(k)} = W^{(1)} + R W^{(k-1)} \tag{11}$$

where

$$W^{(1)} = \frac{\int_0^L P(x) \sin\left(\frac{\pi}{L} x\right) dx}{E^* \pi^4 \left(\frac{b h^3}{24 L^3}\right)} \tag{12}$$

is the amplitude of the local beam deflection.

Using equation (11), the amplitude of Euler-Bernoulli nonlocal beams can be easily determined after a few iterations. In equation (11), $R$ is introduced as a nonlocal residual-based correction factor that corrects the local-like field equation for the nonlocal residual. This correction factor is determined for the considered simple supported beam, as follows:

$$R = 1 - \frac{\left(\frac{\nu}{1-\nu}\right)\left(\epsilon_2 \left(\frac{\pi}{L}\right)^2 + 1\right) + \left(\frac{1-2\nu}{1-\nu}\right)\left(\epsilon_1 \left(\frac{\pi}{L}\right)^2 + 1\right)}{1 + (\epsilon_1 + \epsilon_2)\left(\frac{\pi}{L}\right)^2 + \epsilon_1 \epsilon_2 \left(\frac{\pi}{L}\right)^4} \tag{13}$$

The derived governing equations based on the general nonlocal theory (equations (11)-(13)) can be simply reduced to their corresponding counterparts based on Eringen's nonlocal theory by setting $\epsilon_1 = \epsilon_2 = \xi^2$. Thus, the correction factor according to Eringen's nonlocal theory will have the form:

$$R = \frac{\xi^2 \left(\frac{\pi}{L}\right)^2}{1 + \xi^2 \left(\frac{\pi}{L}\right)^2} \tag{14}$$

As previously demonstrated, the correction factor depends on Poisson's ratio in the context of the general nonlocal theory. Eringen's nonlocal theory does not account for Poisson's ratio on nonlocal residuals of the beam.

## 2.3. Frequencies of simple supported beams

To investigate the free vibration of simple supported Euler-Bernoulli nonlocal beams, the forcing term is eliminated from equation (5) and the deflection and its correction field are decomposed as follows:

$$w(x,t) = W \sin\left(\frac{n\pi}{L}x\right)\exp(\mathrm{i}\omega_n t)$$
$$w_c(x,t) = W_c \sin\left(\frac{n\pi}{L}x\right)\exp(\mathrm{i}\omega_n t)$$

(15)

where $\omega_n$ is the natural frequency of an $n$th mode of beam's vibration.

The equation of motion of the simple supported beam is derived according to the iterative nonlocal residual approach in the following form by substituting equation (15) into, respectively, equations (7), (6), and (5) and integrating the result over the beam length:

$$W^{(k)}\left[1 - \left(\frac{12\rho L^4}{E^* h^2 (n\pi)^4}\right)\omega_n^2\right] = R_n W^{(k-1)}$$

(16)

where

$$R_n = 1 - \frac{\left(\frac{\nu}{1-\nu}\right)\left(\epsilon_2\left(\frac{n\pi}{L}\right)^2 + 1\right) + \left(\frac{1-2\nu}{1-\nu}\right)\left(\epsilon_1\left(\frac{n\pi}{L}\right)^2 + 1\right)}{1 + (\epsilon_1 + \epsilon_2)\left(\frac{n\pi}{L}\right)^2 + \epsilon_1\epsilon_2\left(\frac{n\pi}{L}\right)^4}$$

(17)

where $R_n$ is the correction factor for the $n$th mode.

Consequently, the natural frequencies of the nonlocal beam can be obtained according to equation (16), as follows:

$$\omega_n^2 = \left(\frac{E^* h^2 (n\pi)^4}{12\rho L^4}\right)(1 - R_n)$$

(18)

With no iterations, the nonlocal frequencies can be directly obtained for the different modes of vibration from equation (18) utilizing the correction factors, $R_n$.

## 3. Model validation

In this section, the results of simple supported nonlocal beams as obtained via the proposed nonlocal model are compared with results of the differential and integral nonlocal models available in the literature for $\epsilon_1 = \epsilon_2 = \xi^2$. As shown in Figures 1 and 2, the proposed model reflects the exact same results as those of the integral and differential nonlocal models (Lu et al., 2006; Fernández-Sáez et al., 2016; Tuna and Kirca, 2016). This demonstrates the effectiveness of the proposed model to represent the static bending and free vibration of simple supported nonlocal beams.

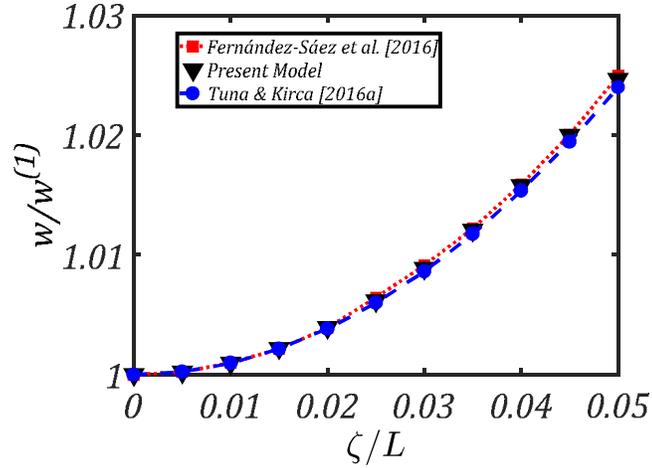

Figure 1: Maximum nonlocal deflection-to-maximum local deflection ratio, $w/w^{(0)}$, of a simple supported beam subjected to a uniform distributed load as a function of the nondimensional nonlocal parameter $\xi/L$.

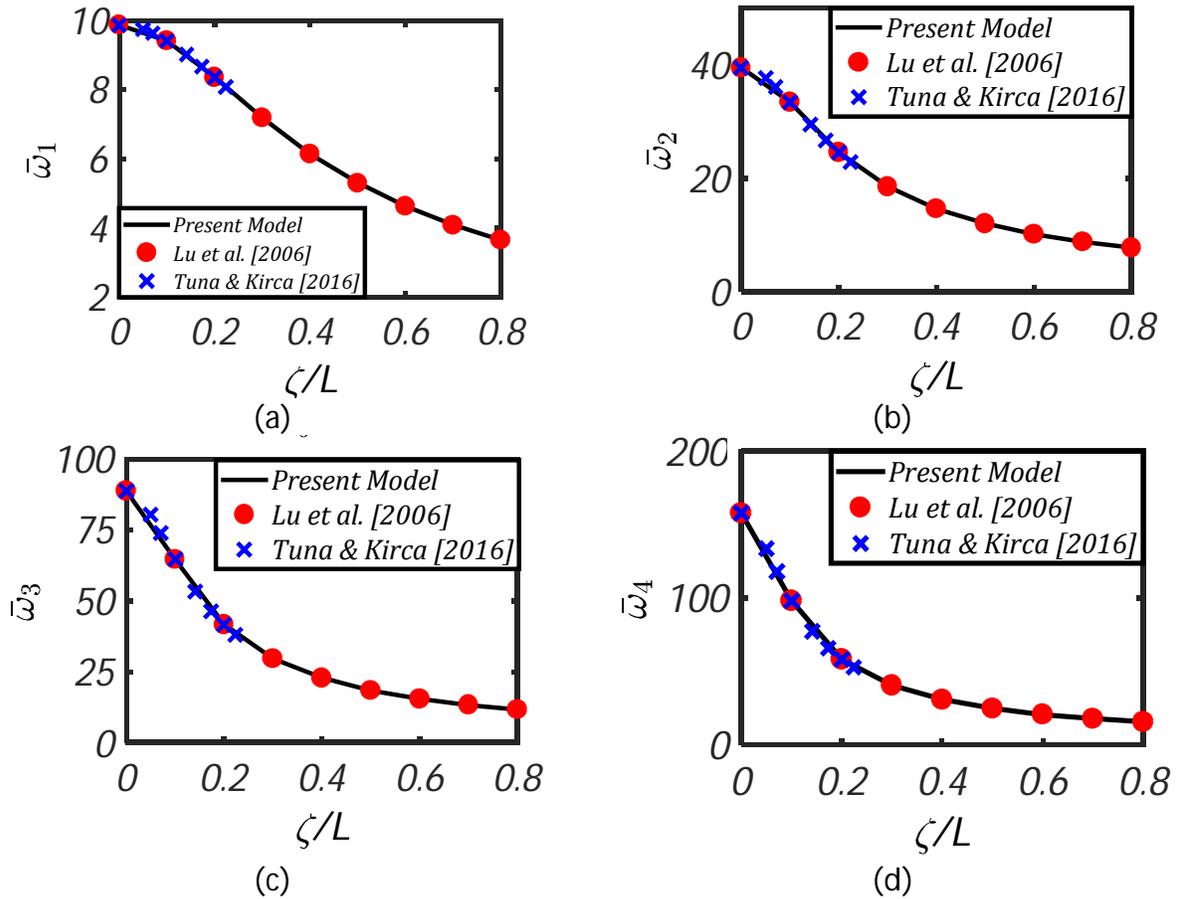

Figure 2: First four nondimensional natural frequencies, $\bar{\omega}_n = \omega_n \sqrt{12\rho L^4/E^* h^2}$, of a simple supported beam as functions of the nondimensional nonlocal parameter $\xi/L$.

## 4. Poisson's ratio effects on mechanics of non-auxetic and auxetic nanobeams

In this section, Poisson's ratio effects on the static bending and free vibration behaviors of non-auxetic and auxetic nanobeams are investigated. To this end, the derived solutions for the simply supported-nonlocal beam based on the iterative-nonlocal residual approach are employed in the next analyses. The dependency of deflections, deformation patterns, and frequencies of nanobeams on the Poisson's ratio are depicted in Figures 3-5 and numerically presented in Tables 1 and 2. In these analyses, the nanobeam is subjected to a nonlocal field such that $\epsilon_2 = 0.4L$ and $\epsilon_1/\epsilon_2 = [0.5,2]$. The following nondimensional parameters are used:

- Nondimensional deflection: $\bar{w}(x) = w(x)(E^*bh^3/12q_0L^4)$ where $q_0$ is the applied load intensity.

- Nondimensional natural frequencies: $\bar{\omega}_n = \omega_n\sqrt{12\rho L^4/E^*h^2}$ where $n$ denotes the mode number.

The nondimensional maximum deflection ($\bar{w}(L/2)$) as a function of the Poisson ratio ($\nu$) is graphically presented in Figure 3 and numerically in Table 1 for different nonlocal parameter ratios ($\epsilon_1/\epsilon_2$). The obtained results demonstrate the significant effects of the Poisson's ratio along with the nonlocal residuals on the nondimensional deflection of nonlocal beams. Thus, depending on the Poisson's ratio and the nonlocal parameters ratio, different behaviors of the beam are depicted. For instance, for the case $\epsilon_1 < \epsilon_2$, the decrease in the Poisson's ratio is accompanied with an increase in the nondimensional beam deflection. On the contrary, when $\epsilon_1 > \epsilon_2$, the beam exhibits a decrease in its nondimensional deflection with the decrease in the Poisson's ratio. This behavior can be attributed to the dependency of the nonlocal field residuals on the Poisson's ratio.

Table 1. Nondimensional maximum deflection, $\bar{w}(x) = w(x)(E^*bh^3/12q_0L^4)$, of simple supported auxetic and non-auxetic nanobeams (uniform distributed load, $p(x) = q_0$).

| $\epsilon_1/\epsilon_2$ | Poisson Ratio ($\nu$) | | | | | | |
|---|---|---|---|---|---|---|---|
| | -1 | -0.75 | -0.5 | -0.25 | 0 | 0.25 | 0.5 |
| 0.5 | 0.0374 | 0.0360 | 0.0342 | 0.0320 | 0.0292 | 0.0254 | 0.0205 |
| 0.75 | 0.0340 | 0.0335 | 0.0329 | 0.0321 | 0.0309 | 0.0292 | 0.0263 |
| 1.0 | 0.0337 | 0.0337 | 0.0337 | 0.0337 | 0.0337 | 0.0337 | 0.0337 |
| 1.25 | 0.0348 | 0.0351 | 0.0356 | 0.0362 | 0.0371 | 0.0388 | 0.0426 |
| 1.5 | 0.0367 | 0.0372 | 0.0380 | 0.0391 | 0.0410 | 0.0444 | 0.0529 |
| 1.75 | 0.0390 | 0.0397 | 0.0408 | 0.0424 | 0.0451 | 0.0504 | 0.0646 |
| 2.0 | 0.0415 | 0.0425 | 0.0438 | 0.0459 | 0.0494 | 0.0565 | 0.0777 |

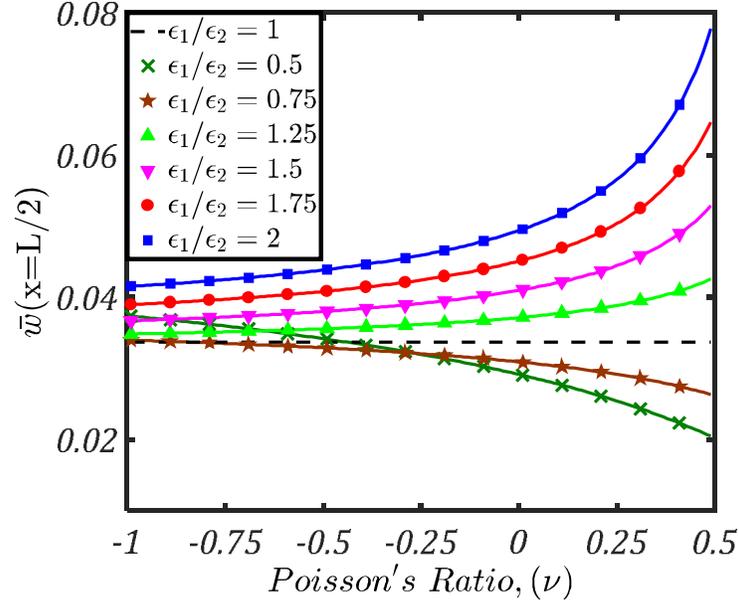

Figure 3: Nondimensional maximum deflection, $\bar{w}(x) = w(x)(E^* b h^3/12 q_0 L^4)$, as a function of the Poisson's ratio, $\nu$, for simple supported auxetic and non-auxetic beams under uniform distributed loads ($p(x) = q_0$). The results are presented for different nonlocal parameter ratios ($\epsilon_1/\epsilon_2$).

It follows from Figure 3 and Table 1 that the general nonlocal theory outperforms Eringen's nonlocal theory in accounting for the Poisson's ratio effects on the deflection of the nonlocal beam. For Eringen's nonlocal theory (which is limit only for the case $\epsilon_1 = \epsilon_2$), the same nondimensional deflection is obtained for the different Poisson's ratio values. The general nonlocal theory, on the contrary, shows the dependency of the nondimensional deflection on the beam's Poisson ratio.

Figure 3 and Table 1 reflect different behaviors for auxetic and non-auxetic beams. Non-auxetic beams ($\nu > 0$) exhibit a specific behavior for the change in the nonlocal fields of the beam's longitudinal and lateral strains. The increase in the nonlocal parameters ratio ($\epsilon_1/\epsilon_2$) is accompanied with an increase in the nondimensional deflection of non-auxetic beams. This behavior demonstrates that nonlocal fields affect non-auxetic beams with a softening mechanism. Auxetic beams ($\nu < 0$), on the other hand, show different behaviors depending on the nonlocal parameters ratio. For $\epsilon_1 < \epsilon_2$, the increase in the nonlocal parameters ratio ($\epsilon_1/\epsilon_2$) is accompanied with a decrease in the nondimensional deflection of auxetic beams. For $\epsilon_1 > \epsilon_2$, the nondimensional deflection of auxetic beams increases with the increase in the nonlocal parameters ratio ($\epsilon_1/\epsilon_2$). Thus, $\epsilon_1 = \epsilon_2$ presents an inflection point at which the contribution of nonlocal fields to the beam behaviors switches from a hardening mechanism (when $\epsilon_1 < \epsilon_2$) to a softening mechanism (when $\epsilon_1 > \epsilon_2$).

Figure 4 shows the distributions of the longitudinal ($\varepsilon_{xx}$) and lateral ($\varepsilon_{yy}$) strains of the beam through its thickness for different Poisson's ratios. Also, the figure represents the deformation patterns of the beam's cross section. The deformation patterns of the beam are shown in Figure 5. It is clear that beams exhibit different strain patterns depending on the Poisson's ratio. For the considered nonlocal field ($\epsilon_1 = 2\epsilon_2$), a decrease in the Poisson's ratio is accompanied with a decrease in both longitudinal and lateral strains of non-auxetic beams (when $\nu > 0$). In contrast, a decrease in the Poisson's ratio gives a decrease in the longitudinal strain and an increase in the lateral strain of auxetic beams (when $\nu < 0$). Moreover, it can be observed that beams exhibit different deformation patterns depending on the Poisson's ratio. The cross section of beams with zero Poisson's ratios ($\nu = 0$) remains the same after deformation (no cross sectional deformations). The cross sections of auxetic and non-auxetic beams ($\nu \neq 0$), in contrast, deform into isosceles trapezoids. This can be attributed to the fact that the upper surface of the beam is subjected to a tension while the lower surface is compressed due to bending. The upper surfaces of non-auxetic beams laterally expand. However, the upper surfaces of auxetic beams laterally shrink. The results presented in figure 4 reveal the significant effect of the Poisson's ratio on the deformation of nanobeams.

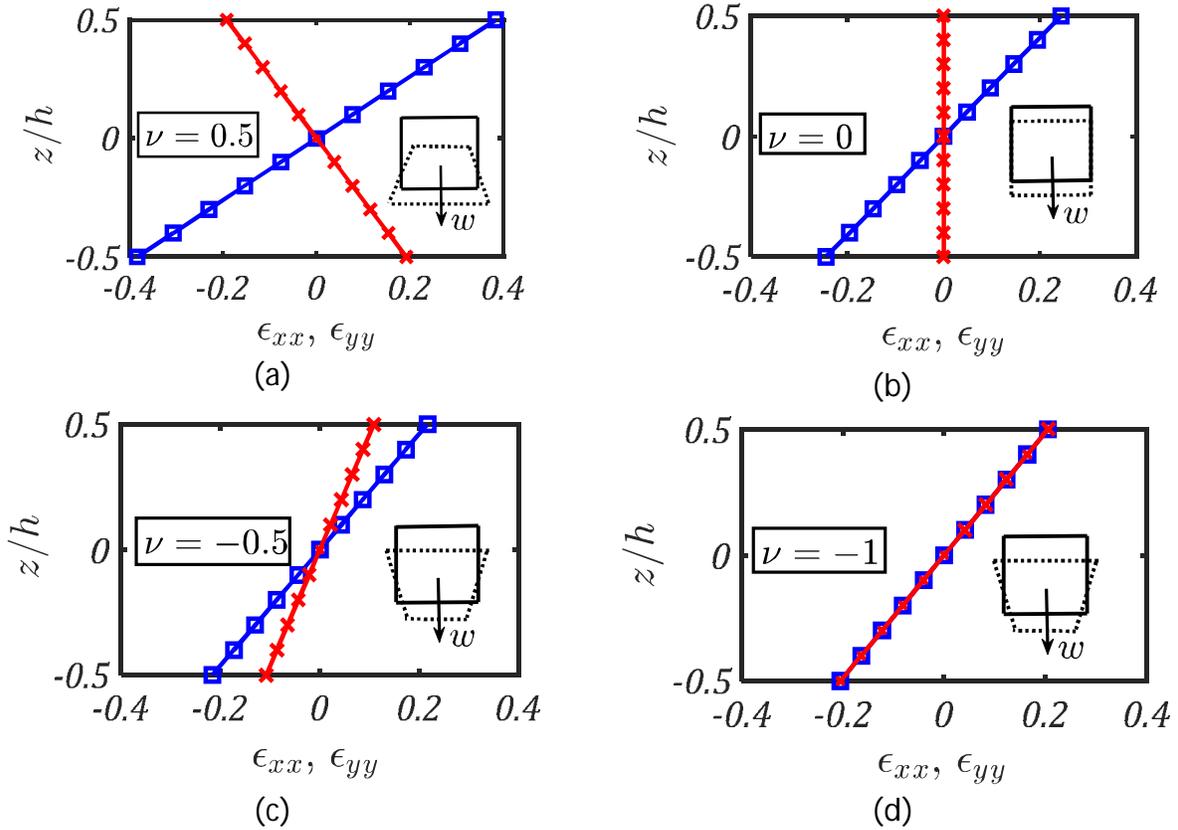

Figure 4: Distributions of longitudinal and lateral strains ($\varepsilon_{xx}$ and $\varepsilon_{yy}$) through the beam thickness (*results are determined for $\epsilon_1 = 2\epsilon_2$*). (a) Non-auxetic beam, (b) beam with zero Poisson's ratio, and (c) and (d) auxetic beams. The longitudinal strain is represented by the blue line while the lateral strain is represented by the red line. The insets in each figure represent the deformed cross-section of the beam.

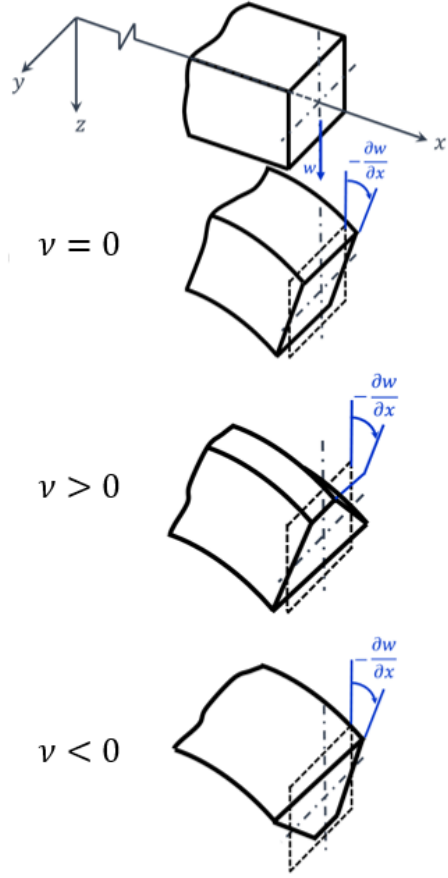

Figure 5: Schematics of deformation patterns in auxetic and non-auxetic beams.

Figure 6 along with Table 2 show the nondimensional natural frequencies $\left( \overline{\omega}_n = \omega_n \sqrt{12 \rho L^4 / E^* h^2} \right)$ of the first four modes as functions of the Poisson ratio ($\nu$) for different nonlocal parameter ratios ($\epsilon_1 / \epsilon_2$). Similar to the beam deflection, different trends are depicted for the natural frequencies for different values of nonlocal parameters. For example, the decrease in the Poisson's ratio is accompanied with a decrease in the beam's nondimensional natural frequencies for beams with $\epsilon_1 < \epsilon_2$. On the other hand, the increase in the nondimensional natural frequencies with the decrease in the Poisson's ratio is observed when $\epsilon_1 > \epsilon_2$. As for the case $\epsilon_1 = \epsilon_2$ (Eringen's nonlocal model), no changes in the nondimensional natural frequencies are obtained when changing the Poisson's ratio.

Table 2: Nondimensional natural frequencies, $\bar{\omega}_n = \omega_n \sqrt{12\rho L^4 / E^* h^2}$, of simple supported auxetic and non-auxetic nanobeams.

| $\epsilon_1 / \epsilon_2$ | Poisson Ratio ($\nu$) | | | | | | |
|---|---|---|---|---|---|---|---|
| | -1 | -0.75 | -0.5 | -0.25 | 0 | 0.25 | 0.5 |
| **1st mode** | | | | | | | |
| 0.5 | 5.8379 | 5.9506 | 6.1028 | 6.3097 | 6.6080 | 7.0773 | 7.8852 |
| 0.75 | 6.1223 | 6.1641 | 6.2213 | 6.3006 | 6.4177 | 6.6081 | 6.9526 |
| 1.0 | 6.1456 | 6.1456 | 6.1456 | 6.1456 | 6.1456 | 6.1456 | 6.1456 |
| 1.25 | 6.0456 | 6.0196 | 5.9835 | 5.9325 | 5.8553 | 5.7243 | 5.4692 |
| 1.5 | 5.8908 | 5.8481 | 5.7885 | 5.7040 | 5.5748 | 5.3526 | 4.9071 |
| 1.75 | 5.7154 | 5.6616 | 5.5863 | 5.4792 | 5.3145 | 5.0281 | 4.4390 |
| 2.0 | 5.5363 | 5.4750 | 5.3892 | 5.2667 | 5.0774 | 4.7451 | 4.0464 |
| **2nd mode** | | | | | | | |
| 0.5 | 14.0418 | 14.6205 | 15.3846 | 16.3945 | 17.8022 | 19.9288 | 23.4134 |
| 0.75 | 14.9572 | 15.1207 | 15.3437 | 15.6506 | 16.1000 | 16.8224 | 18.1041 |
| 1.0 | 14.5951 | 14.5951 | 14.5951 | 14.5951 | 14.5951 | 14.5951 | 14.5951 |
| 1.25 | 13.9484 | 13.8698 | 13.7604 | 13.6058 | 13.3705 | 12.9689 | 12.1770 |
| 1.5 | 13.2732 | 13.1528 | 12.9845 | 12.7451 | 12.3774 | 11.7389 | 10.4317 |
| 1.75 | 12.6407 | 12.4967 | 12.2949 | 12.0067 | 11.5609 | 10.7771 | 9.1207 |
| 2.0 | 12.0673 | 11.9097 | 11.6884 | 11.3714 | 10.8785 | 10.0033 | 8.1032 |
| **3rd mode** | | | | | | | |
| 0.5 | 22.2553 | 23.4171 | 24.9338 | 26.9140 | 29.6372 | 33.6903 | 40.2266 |
| 0.75 | 23.6540 | 23.9450 | 24.3414 | 24.8857 | 25.6806 | 26.9534 | 29.1991 |
| 1.0 | 22.7743 | 22.7743 | 22.7743 | 22.7743 | 22.7743 | 22.7743 | 22.7743 |
| 1.25 | 21.5583 | 21.4294 | 21.2499 | 20.9961 | 20.6095 | 19.9485 | 18.6408 |
| 1.5 | 20.3826 | 20.1887 | 19.9175 | 19.5316 | 18.9379 | 17.9047 | 15.7775 |
| 1.75 | 19.3244 | 19.0954 | 18.7742 | 18.3152 | 17.6043 | 16.3508 | 13.6833 |
| 2.0 | 18.3884 | 18.1400 | 17.7909 | 17.2905 | 16.5113 | 15.1239 | 12.0875 |
| **4th mode** | | | | | | | |
| 0.5 | 30.3836 | 32.1331 | 34.4048 | 37.3537 | 41.3850 | 47.3471 | 56.8996 |
| 0.75 | 32.1969 | 32.6111 | 33.1752 | 33.9490 | 35.0778 | 36.8825 | 40.0596 |
| 1.0 | 30.8121 | 30.8121 | 30.8121 | 30.8121 | 30.8121 | 30.8121 | 30.8121 |
| 1.25 | 29.0546 | 28.8770 | 28.6299 | 28.2802 | 27.7474 | 26.8359 | 25.0302 |
| 1.5 | 27.4026 | 27.1373 | 26.7664 | 26.2382 | 25.4255 | 24.0099 | 21.0889 |
| 1.75 | 25.9369 | 25.6251 | 25.1879 | 24.5626 | 23.5937 | 21.8838 | 18.2351 |
| 2.0 | 24.6519 | 24.3148 | 23.8411 | 23.1616 | 22.1032 | 20.2165 | 16.0751 |

As expected, different trends are obtained for the natural frequencies of the auxetic and non-auxetic beams. The nondimensional natural frequencies are obtained decreasing with the increase in the nonlocal parameter for non-auxetic beams. Thus, nonlocal fields affect non-auxetic beams with a softening mechanism. However, auxetic beams give different trends depending on the residual nonlocal fields of the beam, as observed in Figure 6.

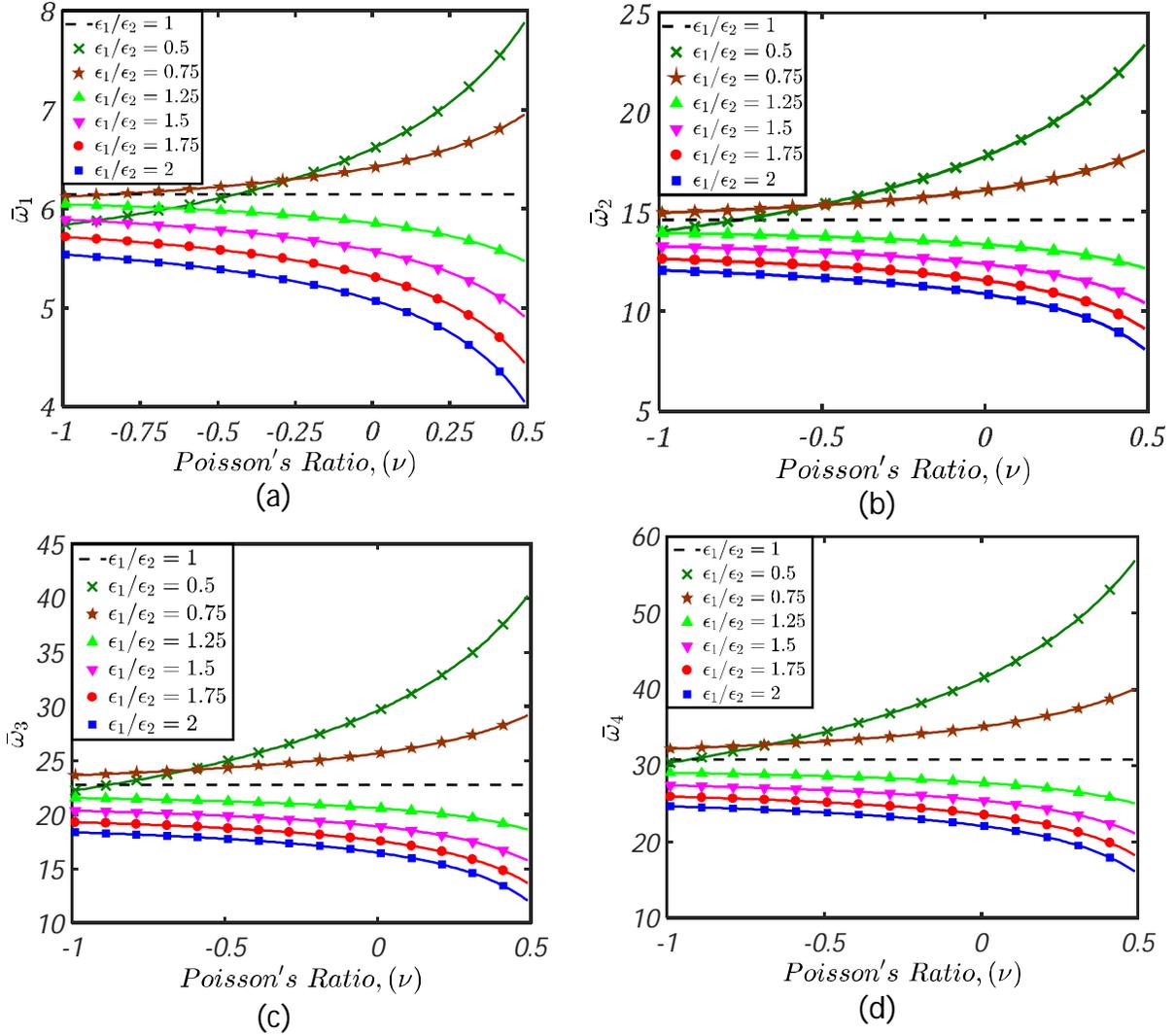

Figure 6: The first four mode-nondimensional natural frequencies, $\bar{\omega}_n = \omega_n\sqrt{12\rho L^4/E^*h^2}$, as functions of Poisson's ratio, $\nu$, for simple supported non-auxetic and auxetic beams. (a) Nondimensional fundamental frequency, (b) nondimensional 2nd-mode natural frequency, (c) nondimensional 3rd-mode natural frequency, and (d) nondimensional 4th-mode natural frequency. Results are presented for different nonlocal parameter ratios ($\epsilon_1/\epsilon_2$).

## Conclusion

Investigations on the effects of Poisson's ratio on the mechanics of auxetic and non-auxetic nanobeams were conducted. The general nonlocal theory and the iterative nonlocal residual approach were employed to develop a parametric study on effects of Poisson's ratio on the static bending and free vibration behaviors of simply-supported auxetic and non-auxetic nanobeams. It was revealed that the static bending as well as natural frequencies of nanobeams strongly depend on the Poisson's ratio and the nonlocal fields of the longitudinal and lateral strains. Moreover, it was demonstrated that non-auxetic beams exhibit increase in deflection and decrease in natural frequencies when increasing nonlocal parameters. Auxetic beams, however, reflect different behaviors depending on the ratio of the nonlocal fields of the longitudinal and lateral strains.

In fact, this study, presents new insights on the mechanics of auxetic beams where the derived results are a benchmark for future work on auxetic beams.